\documentclass{article}
\usepackage{spconf}
\usepackage{cite}
\usepackage{amsmath,amssymb,amsfonts,url,times,booktabs, tabularx,csquotes,hyperref}
\usepackage{algorithmic}
\usepackage{graphicx}
\usepackage{textcomp}
\usepackage{xcolor}
\usepackage{orcidlink}
\usepackage{adjustbox}
\usepackage{multirow,afterpage}
\usepackage{tikz,pgfplots}
\usepackage{amsbsy}  
\usepackage{acronym}
\usepackage{siunitx}
\usepackage{bbm}
\pgfplotsset{compat=1.18}
\usetikzlibrary{shapes.geometric, arrows, calc, fit, shapes, backgrounds}

\acrodef{ai}[AI]{artificial intelligence}
\acrodef{asr}[ASR]{automatic speech recognition}
\acrodef{cnn}[CNN]{convolutional neural network}
\acrodef{ctc}[CTC]{connectionist temporal classification}
\acrodef{dtw}[DTW]{dynamic time warping}
\acrodef{dba}[DBA]{\acs{dtw} barycenter averaging}
\acrodef{hfcc}[HFCC]{human factor cepstral coefficient}
\acrodef{kws}[KWS]{keyword spotting}
\acrodef{mfcc}[MFCC]{Mel-frequency cepstral coefficient}
\acrodef{stft}[STFT]{short-time Fourier transform}
\newcommand{\BE}{\begin{equation*}\begin{aligned}}
\newcommand{\EE}{\end{aligned}\end{equation*}}

\title{TACos: Learning Temporally Structured Embeddings for Few-Shot Keyword Spotting with Dynamic Time Warping}
\name{Kevin Wilkinghoff~$^{1}$ and Alessia Cornaggia-Urrigshardt~$^{1}$}
\address{
$^{1}$Fraunhofer FKIE, Fraunhoferstraße 20, 53343 Wachtberg, Germany \\
kevin.wilkinghoff@ieee.org, alessia.cornaggia-urrigshardt@fkie.fraunhofer.de}
\begin{document}
\ninept
\maketitle
\begin{abstract}
To segment a signal into blocks to be analyzed, few-shot \acl{kws} (\acs{kws}) systems often utilize a sliding window of fixed size.
Because of the varying lengths of different keywords or their spoken instances, choosing the right window size is a problem:
A window should be long enough to contain all necessary information needed to recognize a keyword but a longer window may contain irrelevant information such as multiple words or noise and thus makes it difficult to reliably detect on- and offsets of keywords.
We propose TACos, a novel angular margin loss for deriving two-dimensional embeddings that retain temporal properties of the underlying speech signal.
In experiments conducted on \acs{kws}-DailyTalk, a few-shot \acs{kws} dataset presented in this work, using these embeddings as templates for \acl{dtw} is shown to outperform using other representations or a sliding window and that using time-reversed segments of the keywords during training improves the performance.
\end{abstract}
\section{Introduction}
\Ac{kws} \cite{lopez-espejo2020deep} is the task of detecting spoken instances of a few pre-defined keywords in audio recordings of possibly long duration.
All other audio content should be ignored and thus \ac{kws} is inherently an open-set classification task.
Additionally, for many \ac{kws} applications only very few training samples are available (few-shot classification \cite{wang2021generalizing_few}) and it is important to detect on- and offsets of detected keywords for further (manual) analysis of the content.
Typical \ac{kws} applications are activating voice assistants \cite{schalkwyk2010your,michaely2017keyword}, searching for content in large databases \cite{moyal2013phonetic} or monitoring audio streams such as (radio) communication transmissions \cite{menon2017radio}.
\par
Many state-of-the-art \ac{kws} systems rely on segmenting audio signals and applying a neural network to extract discriminative embeddings for each segment that can be used to detect keywords \cite{kamper2016deep,ma2019hypersphere}.
For few-shot \ac{kws}, neural networks with a prototypical loss \cite{snell2017prototypical} are often used to learn an embedding function \cite{mazumder2021few-shot,parnami2022few,kim2022dummy}.
Similar approaches are used for few-shot detection of bioacoustic events \cite{nolasco2022few-shot} or sound events in general \cite{wang2020few-shot,wang2022active}.
Usually, a sliding window is applied to segment the signal into blocks of fixed size, in which keywords are searched.
The chosen window size needs to be long enough to capture sufficient information for identifying a keyword.
However, longer windows are likely to contain multiple keywords or, in case of a short keyword, too much irrelevant information thus degrading the performance.
Additionally, precisely estimating on- and offsets of detected keywords is much more difficult with longer windows.
Furthermore, the length of different keywords can strongly vary making it difficult to determine a suitable, fixed, window length.
\par
In \ac{asr} \cite{li2022recent}, this problem is solved by using sequence-to-sequence losses such as the \acl{ctc} loss \cite{graves2006connectionist}.
However, training with such losses requires sufficient amounts of data making them unsuitable for a few-shot classification task.
Although it is also possible to use a pre-trained \ac{asr} system \cite{kim2019query,li2021query} or pre-trained \ac{asr} embeddings \cite{kirandevraj2022generalized}, this requires collecting many hours of training data recorded in similar acoustic environments and creates a large computational overhead.
In \cite{jeffares2022spike}, it has been proposed to save computational power by providing the network the ability to spike once a keyword is detected and immediately stop analyzing the remaining part of the input sequence.
Classically, hand-crafted speech features such as \acp{hfcc} \cite{von2010perceptual} are used as two-dimensional templates for \ac{dtw}.
However, the performance of these unsupervised approaches quickly degrades for very short words or in difficult acoustic conditions.
It is also possible to combine multiple \ac{kws} approaches:
In \cite{menon2018fast}, \ac{dtw} and hand-crafted speech features are used to obtain training data for a neural network-based \ac{kws} system.
In our prior work \cite{wilkinghoff2021twodimensional}, two-dimensional embeddings, to be used as features for \ac{dtw}, are trained using a neural network applied to windowed segments of audio signals.
Still, the obtained embeddings are mostly constant over time and thus many problems resulting from using a sliding window persist.
\par
The contributions of this work are the following:
First and foremost, TACos, a novel loss function for learning embeddings that also capture the temporal structure, and a \ac{dtw}-based few-shot \ac{kws} system are proposed.
Second, the few-shot \ac{kws} dataset \ac{kws}-DailyTalk\footnote{\url{https://github.com/wilkinghoff/kws-dailytalk}} based on the \ac{asr} dataset DailyTalk \cite{lee2022dailytalk} is presented.
In contrast to existing \ac{kws} datasets such as SpeechCommands \cite{warden2018speech}, \ac{kws}-DailyTalk is an open-set classification dataset with isolated keywords as training data and complete spoken sentences as validation and test data.
The proposed \ac{kws} system is shown to outperform systems using hand-crafted speech features, a sliding window or other \ac{kws} embeddings.
Furthermore, it is shown that teaching the model to distinguish between regular and temporally reversed segments improves the performance.

\section{Methodology}
\begin{figure*}[tbh!]
	\centering
	\begin{adjustbox}{max width=\textwidth}
		\usetikzlibrary{shapes.geometric, arrows, calc, fit}
\tikzset{box_both/.style={top color=teal!20!red!35,shade,draw, rectangle, rounded corners, thick, node distance=1cm, minimum width=2.3cm, text centered, minimum height=2em, text width=2.8cm}}
\tikzset{box_train/.style={top color=teal!85!blue!85,shade,draw, rectangle, rounded corners, thick, node distance=1cm, minimum width=2.3cm, text centered, minimum height=2em, text width=2.8cm}}
\tikzset{box_inf/.style={top color=teal!15!yellow!30,shade,draw, rectangle, rounded corners, thick, node distance=1cm, minimum width=2.3cm, text centered, minimum height=2em, text width=2.8cm}}
\tikzset{container_both/.style={fill=teal!35!red!50,draw, rectangle, dashed, ultra thick, inner sep=1em,minimum width=2cm}}
\tikzset{container_train/.style={fill=teal!70!blue!70,draw, rectangle, dashed, ultra thick, inner sep=0.5em,minimum width=2cm}}
\tikzset{container_inf/.style={fill=teal!15!yellow!45,draw, rectangle, dashed, ultra thick, inner sep=0.5em,minimum width=2cm}}
\tikzset{font={\fontsize{10pt}{10}\selectfont}}
\tikzstyle{arrow} = [thick, ->, >=stealth]
\begin{tikzpicture}
\node(kw_label)[box_train,node distance=1.65cm]{obtain keyword label};
\node(wav)[box_both,below of=kw_label,node distance=1.5cm]{obtain raw waveform};
\node(prep)[box_both,right of=wav, node distance=4cm]{pre-process waveform};
\node(segs)[box_both,below of=prep, node distance=1.5cm]{divide into segments};
\node(enc)[box_train,right of=prep, node distance=4cm, yshift=1.5cm]{encode positional information};
\node(rev)[box_train,below of=enc, node distance=1.5cm]{revert along temporal axis};
\node(log_mel)[box_both,below of=rev, node distance=1.5cm]{compute magnitude log-Mel spectrograms};
\node(aug)[box_train,right of=enc, node distance=4cm]{apply data augmentation};
\node(train)[box_train,right of=rev, node distance=4cm]{train neural network};
\node(emb)[box_inf,below of=train, node distance=1.5cm]{extract embeddings with neural network};
\node(cost)[box_inf,right of=emb, node distance=4cm]{compute cost matrices};
\node(dtw)[box_inf,above of=cost, node distance=1.5cm]{apply sub-sequence \ac{dtw}};
\node(post)[box_inf,above of=dtw, node distance=1.5cm]{post-process potential matches};
\node(out)[box_inf,right of=dtw, node distance=4cm]{return keyword matches with on- and offsets};

\begin{scope}[on background layer]
\node[container_both, fit=(emb)(aug)(train)] (nn) {};
\node[container_train, fit=(aug)(train)] (nn_train) {};
\node[container_inf, fit=(emb)] (nn_inf) {};
\node[container_both, fit=(prep)(segs)(log_mel)(enc)(rev)] (frontend) {};
\node[container_inf, fit=(cost)(dtw)(post)] (backend_inf) {};
\node[container_train, fit=(enc)(rev)] (frontend_train) {};
\node at (backend_inf.south) [text centered,below,node distance=0 and 0, align=center,xshift=0cm] (backendtxt) {backend};
\node at (nn.south) [text centered,below,node distance=0 and 0, align=center,xshift=0cm] (nntxt) {neural network};
\node at (frontend.south) [text centered,below,node distance=0 and 0, align=center,xshift=0cm] (frontendtxt) {frontend};
\end{scope}

\draw[arrow](wav.0) [out=0, in=180] to (prep.180);
\draw[arrow](kw_label.0) [out=0, in=160] to (aug.160);
\draw[arrow](prep.270) [out=270, in=90] to (segs.90);
\draw[arrow](segs.0) [out=0, in=180] to (log_mel.180);
\draw[arrow](segs.4) [out=35, in=180] to (enc.180);
\draw[arrow](rev.270) [out=270, in=90] to (log_mel.90);
\draw[arrow](segs.2) [out=5, in=180] to (rev.180);
\draw[arrow](rev.90) [out=90, in=270] to (enc.270);
\draw[arrow](log_mel.2) [out=20, in=200] to (aug.200);
\draw[arrow](log_mel.0) [out=0, in=180] to (emb.180);
\draw[arrow](enc.0) [out=0, in=180] to (aug.180);
\draw[arrow](aug.270) [out=270, in=90] to (train.90);
\draw[arrow](emb.0) [out=0, in=180] to (cost.180);
\draw[arrow](cost.90) [out=90, in=270] to (dtw.270);
\draw[arrow](dtw.90) [out=90, in=270] to (post.270);
\draw[arrow](post.0) [out=0, in=180] to (out.180);

\end{tikzpicture}
	\end{adjustbox}
	\caption{Structure of the proposed \ac{kws} system. Blocks colored in blue are only used for training the system, blocks colored in yellow are only used for inference and blocks colored in red are used for training and inference.}
	\label{fig:system_structure}
\end{figure*}
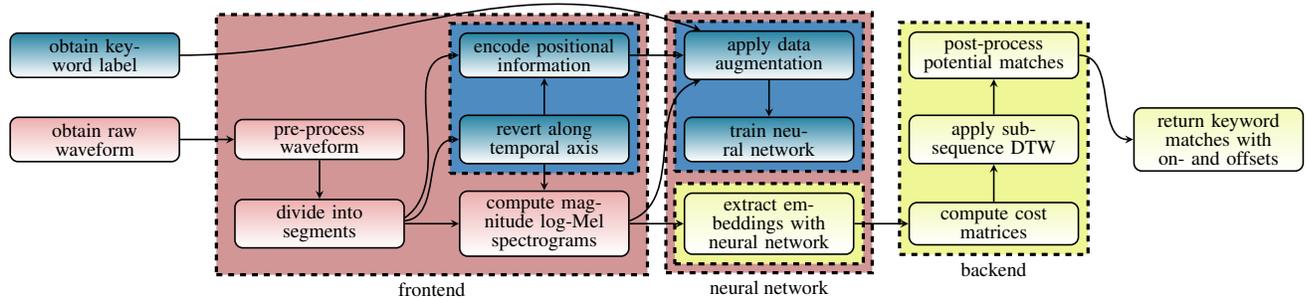
This work is based on prior work on learning two-dimensional embeddings for \ac{kws} with \ac{dtw} \cite{wilkinghoff2021twodimensional}.
A \ac{kws} system using these embeddings can be divided into a frontend for pre-processing the data, a neural network for extracting embeddings, and a backend for computing scores and finding keywords using \ac{dtw}.
To improve the quality of the embeddings, we propose 1) TACos, a novel angular margin loss that also considers the position of a segment in a keyword, and 2) recognizing temporally reversed segments during training.
In the following, each component of the \ac{kws} system will be reviewed and afterwards both proposed improvements will be presented.
An overview of the resulting \ac{kws} system is depicted in \autoref{fig:system_structure}.

\subsection{Review of embeddings for KWS with DTW}
\label{sec:review}
\textbf{Frontend:}
First, all waveforms are converted to single-channel, normalized to an amplitude of $1$, resampled to \SI{16}{\kilo\hertz} and high-pass filtered at \SI{50}{\hertz}.
For the training samples containing isolated keywords, the waveforms are divided into overlapping segments with a length of $L_\text{seg}=\SI{0.25}{\second}$ and an overlap of $\frac{L_\text{seg}}{5}$.
During inference, a segment overlap of $\frac{256}{\SI{16000}{\hertz}}$ is used to increase the temporal resolution of the resulting embeddings.
Furthermore, samples are padded with $\big\lfloor \frac{L_\text{seg}\cdot16000}{2}\big\rfloor$ zeros on both sides to ensure that the centers of the extracted segments align with their temporal position in the audio signal.
Segments shorter than $L_\text{seg}$ are padded with zeros.
From these, log-Mel magnitude spectrograms with $64$ Mel bins are extracted using an \acs{stft} with Hanning-weighted windows of size $1024$ and a hop size of $256$.

\textbf{Neural network:}
\label{sec:nn}
To extract two-dimensional embeddings, the  modified ResNet architecture from \cite{wilkinghoff2021twodimensional} is used.
This model consists of four times two residual blocks \cite{he2016residual}, each using convolutional layers with filters of size $3\times3$, max-pooling for the frequency dimension and dropout with a probability of $20\%$, followed by a global max-pooling operation over the frequency dimension and a dense layer without activation function.
Throughout the network, the same time dimension is kept by padding appropriately and not applying temporal pooling operations.
The model, without the loss function, has only $713,486$ trainable parameters.
As a loss function, the AdaCos loss \cite{zhang2019adacos}, which is an angular margin loss for classification with a dynamically adaptive scale parameter, is used to discriminate between different keywords.
Additional details can be found in \cite{wilkinghoff2021twodimensional}.
\par
The network is trained for $1000$ epochs with a batch size of $32$ using Adam \cite{kingma2015adam}.
Most keywords have different lengths resulting in a class imbalance due to a different total number of segments for each keyword class.
To handle this, random oversampling is applied.
For data augmentation, Mixup \cite{zhang2017mixup} with a mixing coefficient drawn from a uniform distribution and SpecAugment \cite{park2019specaugment} are used.
During training, random segments of the background noise recordings from SpeechCommands \cite{warden2018speech} are used as an additional \enquote{no speech} class.

\textbf{Backend:}
\label{sec:backend}
The backend consists of applying sub-sequence \ac{dtw} \cite{kurth2010analysis}.
All embeddings belonging to different segments of the same audio signal are combined by taking the mean of all individual frames of the time-frequency representation that overlap in time resulting in \ac{dtw} templates.
In a next step, cost matrices are computed by applying the pairwise cosine distance between the templates of test sentences and the templates extracted from individual training samples.
We also experimented with computing Fr\'echet means of all templates belonging to the same keyword with the \ac{dba} algorithm \cite{petitjean2011global} but this led to worse performance than using the templates of the individual training samples.
To compute accumulated cost matrices, the \ac{dtw} step sizes $(2,1)$, $(1,1)$ and $(1,2)$ are used.
Note that computing the accumulated cost matrices can be parallelized by sweeping diagonally over the cost matrix.
In total, the presented \ac{kws} approach is much faster than real-time.
For each temporal position, a warping path is calculated and the corresponding accumulated cost is normalized with respect to the path length.
The negative accumulated costs serve as matching scores that can be compared to a pre-defined threshold.
Scores exceeding the threshold are considered as valid detections of a keyword and the start and end positions of the corresponding paths are returned as on- and offsets, respectively.
If multiple detections overlap in time, all detections are shortened such that, at each position in time, only the single detection with the highest score is kept.
Detections with less than half the duration of the training sample belonging to the detected keyword are discarded.

\subsection{TACos loss function}

\begin{figure}[tbh!]
	\centering
	\begin{adjustbox}{max width=\columnwidth}
  \includegraphics{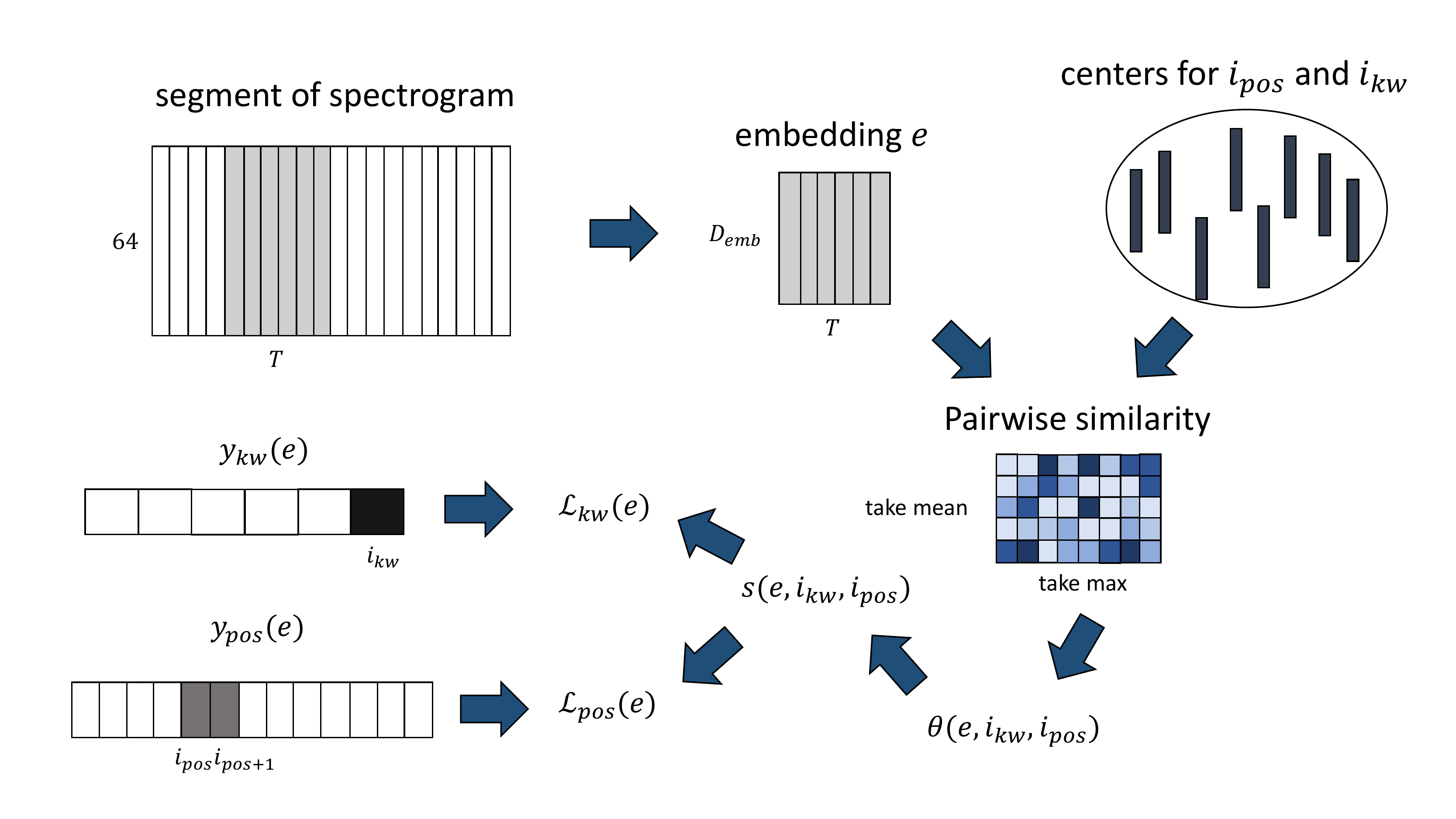}
	\end{adjustbox}
     \caption{Illustration of the TACos loss function.}
	\label{fig:tacos}
\end{figure}
The TACos loss function, illustated in \autoref{fig:tacos}, consists of a supervised loss $\mathcal{L}_{\text{kw}}$ for predicting the keyword a given audio segment belongs to, which is the same loss as defined in \cite{wilkinghoff2021twodimensional}, and a self-supervised loss $\mathcal{L}_{\text{pos}}$ for predicting the relative position of this segment within a keyword.
When only using a supervised loss, the resulting embeddings are mostly constant over time.
The idea of introducing a positional loss is to force the network to learn two-dimensional embeddings that are changing over time and thus are more suitable to be used as templates for \ac{dtw}. 
\par
Since the length of different keywords may vary substantially, the absolute position of a segment within a keyword has to be encoded relative to the length of the keyword to be able to use a fixed number of position classes for all keywords.
Let $K\in\mathbb{N}$ be the total number of training samples, $N_\text{seg}(k)\in\mathbb{N}$ denote the number of segments belonging to the $k$th training sample and define ${N_\text{pos}:=\max_{k\in\lbrace1,...,K\rbrace}\lbrace N_\text{seg}(k)\rbrace}$.
Then, the relative positional encoding ${y_\text{pos}(e_{k,i_\text{seg}})\in[0,1]^{N_\text{pos}}}$ of embedding $e_{k,i_\text{seg}}$ belonging to segment $i_\text{seg}\in\lbrace1,...,N_\text{seg}(k)\rbrace$ of keyword sample $k$ is obtained by setting
\BE y_\text{pos}(e_{k,i_\text{seg}},i_\text{pos})=\frac{\mathbbm{1}_{I_\text{active}(k,i_\text{seg})}(i_\text{pos})}{\sum_{j_\text{pos}=1}^{N\text{pos}}\mathbbm{1}_{I_\text{active}(k,i_\text{seg})}(j_\text{pos})}\EE
with $I_\text{active}(k,i_\text{seg})=\big[1+\big\lceil \frac{(i_\text{seg}-1)\cdot N_\text{pos}}{N_\text{seg}(k)}\big\rceil,\big\lceil \frac{i_\text{seg}\cdot N_\text{pos}}{N_\text{seg}(k)}\big\rceil\big]$.
Thus, for keyword samples shorter than the longest training sample, multiple positions may be set as active with equal probability resulting in soft labels for the position.
\par
Let $N_\text{kw}\in\mathbb{N}$ denote the number of different keywords in the dataset and let $y_\text{kw}(e,i_\text{kw})\in[0,1]$ denote the categorical keyword labels of the samples.
Let ${T:=\big\lceil L_\text{seg}\cdot\frac{16000}{256}\big\rceil\in\mathbb{N}}$ be the time dimension and ${D_\text{emb}\in\mathbb{N}}$ be the embedding dimension of an embedding ${e\in\mathbb{R}^{T\times D_\text{emb}}}$ belonging to a single segment.
Then, we define the cosine similarity of embedding $e$ to keyword ${i_\text{kw}\in\lbrace1,...,N_\text{kw}\rbrace}$ and position ${i_\text{pos}\in\lbrace1,...,N_\text{pos}\rbrace}$ as a temporal mean given by
\BE\label{eq:theta} \theta(e,i_\text{kw},i_\text{pos}):=&\frac{1}{T}\sum_{t=1}^T\max_{i_\text{cluster}}\frac{\langle e(t),c(i_\text{cluster},i_\text{kw},i_\text{pos})\rangle}{\lVert e(t)\rVert_2\lVert c(i_\text{cluster},i_\text{kw},i_\text{pos})\rVert_2}\EE
for trainable cluster centers $c\in\mathbb{R}^{N_\text{cluster}\times N_\text{kw} \times N_\text{pos}\times D_\text{emb}}$ with $N_\text{cluster}\in\mathbb{N}$, which do not have a temporal dimension.
The corresponding softmax probability of embedding $e$ belonging to keyword $i_\text{kw}$ and position $i_\text{pos}$ is defined as
\BE s(e,i_\text{kw},i_\text{pos}):=&\frac{\exp(\hat{s}\cdot \theta(e,i_\text{kw},i_\text{pos}))}{\sum_{j_{\text{kw}=1}}^{N_{\text{kw}}}\sum_{j_\text{pos}=1}^{N_\text{pos}}\exp(\hat{s}\cdot \theta(e,j_\text{kw},j_\text{pos}))}\EE
where $\hat{s}\in\mathbb{R}_+$ is the dynamically adaptive scale parameter as defined for the sub-cluster AdaCos loss in \cite{wilkinghoff2021sub}.
The probability of embedding $e$ belonging to keyword $i_\text{kw}$ is set to $\sum_{j_\text{pos}=1}^{N_\text{pos}}s(e,i_\text{kw},j_\text{pos})$ and the probability of embedding $e$ belonging to position $i_\text{pos}$ is set to $\sum_{j_\text{kw}=1}^{N_\text{kw}}s(e,j_\text{kw},i_\text{pos})$.
Therefore, the loss functions for a single embedding $e$ are equal to
\BE \mathcal{L}_\text{kw}(e)&:=\sum_{i_\text{kw}=1}^{N_\text{kw}}y_\text{kw}(e,i_\text{kw})\log\big(\sum_{i_\text{pos}=1}^{N_\text{pos}}s(e,i_\text{kw},i_\text{pos})\big)\\
\mathcal{L}_\text{pos}(e)&:=\sum_{i_\text{pos}=1}^{N_\text{pos}}y_\text{pos}(e,i_\text{pos})\log\big(\sum_{i_\text{kw}=1}^{N_\text{kw}}s(e,i_\text{kw},i_\text{pos})\big)\EE
and the TACos loss used for training the network is
\BE \mathcal{L}_\text{TAC}&:=-\frac{1}{K}\sum_{k=1}^K\frac{1}{N_\text{seg}(k)}\sum_{i_\text{seg}=1}^{N_\text{seg}(k)}(\mathcal{L}_\text{kw}(e_{k,i_\text{seg}})+\mathcal{L}_\text{pos}(e_{k,i_\text{seg}})).\EE
Note that the cluster centers $c$ significantly increase the total number of trainable parameters of the model.
For all experiments in this work, ${D_\text{emb}=128}$ and ${N_\text{cluster}=16}$ were used.

\subsection{Using temporally reversed segments}
As a second modification, we propose to utilize temporally reversed versions of all keyword segments as additional training samples when training the embedding model.
The idea is that the network has to be able to recognize the correct temporal order, i.e., we enforce that the reverse keyword is considered to be different from the non-reversed version.
By doing so, the model has a harder task in correctly predicting the corresponding keyword and position of the segments.
This leads to more informative embeddings and reduces the number of false detections.
For each keyword class except for \enquote{no speech}, an additional unique label for the reversed keyword segments is introduced, almost doubling the number of different keyword classes.
The position of the reversed segments and the segments not containing any speech are both encoded by using a uniform posterior probability over the position classes.

\section{Experiments}
\begin{table*}[tbh!]
	\centering
	\caption{Event-based, micro-averaged F-score, precision and recall obtained on \ac{kws}-DailyTalk with different \ac{kws} systems. Highest F-scores for each feature representation are highlighted with bold letters, overall highest F-scores are underlined.}
\begin{adjustbox}{max width=\textwidth}
	\begin{tabular}{llllll|lll}
		\toprule
		\ac{kws} feature representation&reversed segments&threshold&\multicolumn{6}{c}{obtained performance}\\
        &&&\multicolumn{3}{c}{validation set}&\multicolumn{3}{c}{test set}\\
        &&&F-score&precision&recall&F-score&precision&recall\\
		\midrule
        \acp{hfcc}&not applicable&global&60.52\%&63.25\%&58.01\%&56.97\%&61.54\%&53.04\% \\
        \acp{hfcc}&not applicable&individual&\pmb{64.71\%}&69.18\%&60.77\%&\pmb{57.74\%}&62.58\%&53.59\%\\
		\midrule
        embeddings (sliding)&not used&global&39.76\%&43.71\%&36.46\%&38.35\%&41.14\%&35.91\%\\
        embeddings (sliding)&not used&individual&46.96\%&49.39\%&44.75\%&40.44\%&40.00\%&40.88\%\\
        embeddings (sliding)&used&global&44.13\%&62.00\%&34.25\%&44.83\%&59.63\%&35.91\%\\
        embeddings (sliding)&used&individual&\pmb{55.43\%}&54.55\%&56.35\%&\pmb{50.42\%}&51.14\%&49.72\%\\
		\midrule
        embeddings ($\mathcal{L}_\text{kw}$) \cite{wilkinghoff2021twodimensional}&not used&global&56.04\%&52.40\%&60.22\%&54.25\%&53.80\%&54.70\%\\
        embeddings ($\mathcal{L}_\text{kw}$) \cite{wilkinghoff2021twodimensional}&not used&individual&58.38\%&57.14\%&59.67\%&53.04\%&53.04\%&53.04\%\\
        embeddings ($\mathcal{L}_\text{kw}$)&used&global&64.58\%&74.64\%&56.91\%&\pmb{61.30\%}&69.72\%&54.70\%\\
        embeddings ($\mathcal{L}_\text{kw}$)&used&individual&\pmb{66.12\%}&64.89\%&67.40\%&60.53\%&57.79\%&63.54\%\\
		\midrule
        embeddings ($\mathcal{L}_\text{kw}+\mathcal{L}_\text{pos}$)&not used&global&62.78\%&75.78\%&53.59\%&64.65\%&82.76\%&53.04\%\\
        embeddings ($\mathcal{L}_\text{kw}+\mathcal{L}_\text{pos}$)&not used&individual&63.36\%&63.19\%&63.54\%&63.31\%&68.15\%&59.12\%\\
        embeddings ($\mathcal{L}_\text{kw}+\mathcal{L}_\text{pos}$)&used&global&65.78\%&82.50\%&54.70\%&\underline{\pmb{70.47\%}}&89.74\%&58.01\%\\
        embeddings ($\mathcal{L}_\text{kw}+\mathcal{L}_\text{pos}$)&used&individual&\underline{\pmb{69.44\%}}&75.00\%&64.64\%&69.16\%&79.29\%&61.33\%\\
		\bottomrule
	\end{tabular}
\end{adjustbox}
\label{tab:performance}
\end{table*}
\subsection{Dataset}
For all experimental evaluations, \ac{kws}-DailyTalk based on the \ac{asr} dataset DailyTalk \cite{lee2022dailytalk} was used.
\ac{kws}-DailyTalk is a five-shot \ac{kws} dataset aimed at detecting $15$ different keywords, namely \enquote{afternoon}, \enquote{airport}, \enquote{cash}, \enquote{credit card}, \enquote{deposit}, \enquote{dollar}, \enquote{evening}, \enquote{expensive}, \enquote{house}, \enquote{information}, \enquote{money}, \enquote{morning}, \enquote{night}, \enquote{visa} and \enquote{yuan}.
The dataset consists of a training set with five isolated samples for each keyword and a duration of 39 seconds, as well as a validation and a test set, with an approximate duration of 10 minutes each, containing $156$ and $157$ sentences taken from dialogues, respectively.
These sentences contain either none, a single, or multiple occurrences of the keywords as well as several other words that are not of interest but may cause false alarms.
All keywords appear about $12$ times each in the validation and the test set.
The on- and offset of each keyword were manually annotated.
Furthermore, it is ensured that a keyword sample used for training and the sentences of the validation and test set that also contain this keyword do not belong to the same conversation to make the \ac{kws} task more realistic and slightly more difficult.
For all experimental evaluations, the event-based F-score (micro-averaged) as implemented in the \emph{sed\_eval} toolbox \cite{mesaros2016metrics} was used.
All hyperparameters of the \ac{kws} systems were tuned to maximize the performance on the validation set.

\subsection{Baseline approaches}
For comparison, the embeddings from \cite{wilkinghoff2021twodimensional} reviewed in \autoref{sec:review} and the following two baseline systems are used.
\par
\textbf{\acp{hfcc}:} 
Instead of applying sub-sequence \ac{dtw} to trained embeddings, \acp{hfcc} \cite{von2010perceptual,kurth2010analysis} based on spectrograms with a window of $\SI{40}{\milli\second}$ and a step size of $\SI{10}{\milli\second}$ are used.
These features were shown to outperform \aclp{mfcc} in query-by-example \Ac{kws} approaches.
The \ac{dtw} algorithm is the same as stated in \autoref{sec:backend}.
\par
\textbf{Sliding window:} As a third system, a sliding window based approach using a neural network trained with the standard AdaCos loss \cite{zhang2019adacos} is used.
The network architecture is the same as presented in \autoref{sec:nn} with the following modifications:
For each residual block, the max-pooling operation is also applied to the temporal dimension and a flattening operation is applied before projecting onto the embedding space.
To detect on- and offsets of keywords, the cosine distance of the resulting embeddings to each of the keyword-specific centers is calculated first.
Then, these cosine distances are compared to a pre-defined threshold and a median filter with a size equal to the average number of segments over all training samples belonging to the corresponding keyword, rounded to the nearest odd natural number, is applied to the Boolean decision results.
Start and end points of the resulting positive regions are adjusted by adding $-\frac{L_\text{seg}}{2}$ and $+\frac{L_\text{seg}}{2}$, respectively, and indicate on- and offsets of detected keywords.

\subsection{Experimental results}
The experimental results obtained on \ac{kws}-DailyTalk can be found in \autoref{tab:performance}.
First, it can be seen that using the proposed TACos loss leads to significantly better performance than when only using $\mathcal{L}_\text{kw}$, as done in \cite{wilkinghoff2021twodimensional}, or using a sliding window based approach, particularly so on the test set.
Moreover, both embeddings perform better than \acp{hfcc} despite having only $39$ seconds of audio recordings available for training.
A second major observation is that using temporally reversed segments for training the embedding model always improves the performance regardless of the chosen \ac{kws} approach.
We want to emphasize that this can be observed although very powerful data augmentation techniques, namely Mixup and SpecAugment, are applied.
Furthermore, tuning individual decision thresholds for each keyword class only improves the test performance when using a sliding window but not for the proposed approach for which the performance actually slightly decreases.
Thus, another advantage is that one does not have to tune individual decision thresholds, which would be very impractical.

\section{Conclusions}
In this paper, TACos, a novel loss function for training neural networks to extract discriminative embeddings with temporal structure and a few-shot \ac{kws}-system based on \ac{dtw} utilizing this loss were proposed.
TACos consists of two components for learning the corresponding keyword of small speech segments as well as their relative position within a given keyword.
To evaluate the performance of the \ac{kws} system, \ac{kws}-DailyTalk, an open-source dataset for few-shot keyword spotting based on DailyTalk, was presented.
In experiments conducted on this dataset, it was shown that the proposed approach outperforms \ac{kws} systems based on other representations or a system using a sliding window.
Last but not least, it was shown that exploiting temporally reversed segments of provided training samples improves the performance regardless of the embedding type.
For future work, the proposed method will be evaluated in noisy conditions and for zero-shot \ac{kws} by pre-training the embeddings on a large \ac{asr} dataset as done in \cite{kirandevraj2022generalized} and/or using pre-trained \ac{asr} embeddings instead of spectrograms as input representations.

\section{Acknowledgments}
The authors would like to thank Paul M. Baggenstoss, Fabian Fritz, Lukas Henneke and Frank Kurth for their valuable feedback.

\bibliographystyle{IEEEbib}
\bibliography{mybib}

\begin{thebibliography}{10}

\bibitem{lopez-espejo2020deep}
Iv{\'{a}}n L{\'{o}}pez{-}Espejo, Zheng{-}Hua Tan, John H.~L. Hansen, and Jesper
  Jensen,
\newblock ``Deep spoken keyword spotting: An overview,''
\newblock {\em {IEEE} Access}, vol. 10, pp. 4169--4199, 2022.

\bibitem{wang2021generalizing_few}
Yaqing Wang, Quanming Yao, James~T. Kwok, and Lionel~M. Ni,
\newblock ``Generalizing from a few examples: {A} survey on few-shot
  learning,''
\newblock {\em {ACM} Comput. Surv.}, vol. 53, no. 3, pp. 63:1--63:34, 2021.

\bibitem{schalkwyk2010your}
Johan Schalkwyk, Doug Beeferman, Fran{\c{c}}oise Beaufays, Bill Byrne, Ciprian
  Chelba, Mike Cohen, Maryam Kamvar, and Brian Strope,
\newblock ``“{Y}our word is my command”: Google search by voice: {A} case
  study,''
\newblock {\em Advances in Speech Recognition: Mobile Environments, Call
  Centers and Clinics}, pp. 61--90, 2010.

\bibitem{michaely2017keyword}
Assaf~Hurwitz Michaely, Xuedong Zhang, Gabor Simko, Carolina Parada, and
  Petar~S. Aleksic,
\newblock ``Keyword spotting for google assistant using contextual speech
  recognition,''
\newblock in {\em ASRU}. 2017, pp. 272--278, {IEEE}.

\bibitem{moyal2013phonetic}
Ami Moyal, Vered Aharonson, Ella Tetariy, and Michal Gishri,
\newblock {\em Phonetic Search Methods for Large Speech Databases},
\newblock Springer Briefs in Electrical and Computer Engineering. Springer,
  2013.

\bibitem{menon2017radio}
Raghav Menon, Armin Saeb, Hugh Cameron, William Kibira, John~A. Quinn, and
  Thomas Niesler,
\newblock ``Radio-browsing for developmental monitoring in {U}ganda,''
\newblock in {\em ICASSP}. 2017, pp. 5795--5799, {IEEE}.

\bibitem{kamper2016deep}
Herman Kamper, Weiran Wang, and Karen Livescu,
\newblock ``Deep convolutional acoustic word embeddings using word-pair side
  information,''
\newblock in {\em ICASSP}. 2016, pp. 4950--4954, {IEEE}.

\bibitem{ma2019hypersphere}
Haoxin Ma, Ye~Bai, Jiangyan Yi, and Jianhua Tao,
\newblock ``Hypersphere embedding and additive margin for query-by-example
  keyword spotting,''
\newblock in {\em APSIPA ASC}. 2019, pp. 868--872, {IEEE}.

\bibitem{snell2017prototypical}
Jake Snell, Kevin Swersky, and Richard~S. Zemel,
\newblock ``Prototypical networks for few-shot learning,''
\newblock in {\em NeurIPS}, 2017, pp. 4077--4087.

\bibitem{mazumder2021few-shot}
Mark Mazumder, Colby~R. Banbury, Josh Meyer, Pete Warden, and Vijay~Janapa
  Reddi,
\newblock ``Few-shot keyword spotting in any language,''
\newblock in {\em Interspeech}. 2021, pp. 4214--4218, {ISCA}.

\bibitem{parnami2022few}
Archit Parnami and Minwoo Lee,
\newblock ``Few-shot keyword spotting with prototypical networks,''
\newblock in {\em {ICMLT}}. 2022, pp. 277--283, {ACM}.

\bibitem{kim2022dummy}
Byeonggeun Kim, Seunghan Yang, Inseop Chung, and Simyung Chang,
\newblock ``Dummy prototypical networks for few-shot open-set keyword
  spotting,''
\newblock in {\em Interspeech}. 2022, pp. 4621--4625, {ISCA}.

\bibitem{nolasco2022few-shot}
In{\^{e}}s Nolasco, Shubhr Singh, E.~Vida{\~{n}}a{-}Villa, E.~Grout,
  J.~Morford, M.~G. Emmerson, F.~H. Jensen, Ivan Kiskin, H.~Whitehead, Ariana
  Strandburg{-}Peshkin, Lisa~F. Gill, Hanna Pamula, Vincent Lostanlen, Veronica
  Morfi, and Dan Stowell,
\newblock ``Few-shot bioacoustic event detection at the {DCASE} 2022
  challenge,''
\newblock in {\em DCASE}, 2022, pp. 136--140.

\bibitem{wang2020few-shot}
Yu~Wang, Justin Salamon, Nicholas~J. Bryan, and Juan~Pablo Bello,
\newblock ``Few-shot sound event detection,''
\newblock in {\em ICASSP}. 2020, pp. 81--85, {IEEE}.

\bibitem{wang2022active}
Yu~Wang, Mark Cartwright, and Juan~Pablo Bello,
\newblock ``Active few-shot learning for sound event detection,''
\newblock in {\em Interspeech}. 2022, pp. 1551--1555, {ISCA}.

\bibitem{li2022recent}
Jinyu Li et~al.,
\newblock ``Recent advances in end-to-end automatic speech recognition,''
\newblock {\em APSIPA Transactions on Signal and Information Processing}, vol.
  11, no. 1, 2022.

\bibitem{graves2006connectionist}
Alex Graves, Santiago Fern{\'{a}}ndez, Faustino~J. Gomez, and J{\"{u}}rgen
  Schmidhuber,
\newblock ``Connectionist temporal classification: labelling unsegmented
  sequence data with recurrent neural networks,''
\newblock in {\em ICML}. 2006, pp. 369--376, {ACM}.

\bibitem{kim2019query}
Byeonggeun Kim, Mingu Lee, Jinkyu Lee, Yeonseok Kim, and Kyuwoong Hwang,
\newblock ``Query-by-example on-device keyword spotting,''
\newblock in {\em ASRU}. 2019, pp. 532--538, {IEEE}.

\bibitem{li2021query}
Li~Meirong, Zhang Shaoying, Cheng Chuanxu, and Xu~Wen,
\newblock ``Query-by-example on-device keyword spotting using convolutional
  recurrent neural network and connectionist temporal classification,''
\newblock in {\em ICSP}, 2021, pp. 1291--1294.

\bibitem{kirandevraj2022generalized}
R.~Kirandevraj, Vinod~Kumar Kurmi, Vinay~P. Namboodiri, and C.~V. Jawahar,
\newblock ``Generalized keyword spotting using {ASR} embeddings,''
\newblock in {\em Interspeech}. 2022, pp. 126--130, {ISCA}.

\bibitem{jeffares2022spike}
Alan Jeffares, Qinghai Guo, Pontus Stenetorp, and Timoleon Moraitis,
\newblock ``Spike-inspired rank coding for fast and accurate recurrent neural
  networks,''
\newblock in {\em ICLR}. 2022, OpenReview.net.

\bibitem{von2010perceptual}
Dirk Von~Zeddelmann, Frank Kurth, and Meinard M{\"u}ller,
\newblock ``Perceptual audio features for unsupervised key-phrase detection,''
\newblock in {\em ICASSP}. IEEE, 2010, pp. 257--260.

\bibitem{menon2018fast}
Raghav Menon, Herman Kamper, John~A. Quinn, and Thomas Niesler,
\newblock ``Fast {ASR}-free and almost zero-resource keyword spotting using
  {DTW} and {CNN}s for humanitarian monitoring,''
\newblock in {\em Interspeech}. 2018, pp. 2608--2612, {ISCA}.

\bibitem{wilkinghoff2021twodimensional}
Kevin Wilkinghoff, Alessia Cornaggia-Urrigshardt, and Fahrettin G\"{o}kg\"{o}z,
\newblock ``Two-dimensional embeddings for low-resource keyword spotting based
  on dynamic time warping,''
\newblock in {\em ITG Speech}. 2021, pp. 9--13, VDE-Verlag.

\bibitem{lee2022dailytalk}
Keon Lee, Kyumin Park, and Daeyoung Kim,
\newblock ``Daily{T}alk: Spoken dialogue dataset for conversational
  text-to-speech,''
\newblock in {\em ICASSP}. 2023, IEEE.

\bibitem{warden2018speech}
Pete Warden,
\newblock ``Speech commands: {A} dataset for limited-vocabulary speech
  recognition,''
\newblock {\em CoRR}, vol. abs/1804.03209, 2018.

\bibitem{he2016residual}
Kaiming He, Xiangyu Zhang, Shaoqing Ren, and Jian Sun,
\newblock ``Deep residual learning for image recognition,''
\newblock in {\em CVPR}. 2016, pp. 770--778, {IEEE}.

\bibitem{zhang2019adacos}
Xiao Zhang, Rui Zhao, Yu~Qiao, Xiaogang Wang, and Hongsheng Li,
\newblock ``Ada{C}os: Adaptively scaling cosine logits for effectively learning
  deep face representations,''
\newblock in {\em CVPR}. 2019, pp. 10823--10832, IEEE.

\bibitem{kingma2015adam}
Diederik~P. Kingma and Jimmy Ba,
\newblock ``Adam: {A} method for stochastic optimization,''
\newblock in {\em ICLR}, 2015.

\bibitem{zhang2017mixup}
Hongyi Zhang, Moustapha Cisse, Yann~N. Dauphin, and David Lopez-Paz,
\newblock ``Mixup: Beyond empirical risk minimization,''
\newblock in {\em ICLR}, 2018.

\bibitem{park2019specaugment}
Daniel~S. Park, William Chan, Yu~Zhang, Chung{-}Cheng Chiu, Barret Zoph,
  Ekin~D. Cubuk, and Quoc~V. Le,
\newblock ``Spec{A}ugment: {A} simple data augmentation method for automatic
  speech recognition,''
\newblock in {\em Interspeech}. 2019, pp. 2613--2617, {ISCA}.

\bibitem{kurth2010analysis}
Frank Kurth and Dirk von Zeddelmann,
\newblock ``An analysis of {MFCC}-like parametric audio features for keyphrase
  spotting applications,''
\newblock in {\em ITG Speech}. 2010, VDE.

\bibitem{petitjean2011global}
Fran{\c{c}}ois Petitjean, Alain Ketterlin, and Pierre Gan{\c{c}}arski,
\newblock ``A global averaging method for dynamic time warping, with
  applications to clustering,''
\newblock {\em Pattern recognition}, vol. 44, no. 3, pp. 678--693, 2011.

\bibitem{wilkinghoff2021sub}
Kevin Wilkinghoff,
\newblock ``Sub-cluster {A}da{C}os: Learning representations for anomalous
  sound detection,''
\newblock in {\em IJCNN}. 2021, IEEE.

\bibitem{mesaros2016metrics}
Annamaria Mesaros, Toni Heittola, and Tuomas Virtanen,
\newblock ``Metrics for polyphonic sound event detection,''
\newblock {\em Applied Sciences}, vol. 6, no. 6, pp. 162, 2016.

\end{thebibliography}

\end{document}